\documentclass[twocolumn,showpacs,preprintnumbers,amsmath,amssymb,prb]{revtex4}

\usepackage{graphicx}
\usepackage{dcolumn}
\usepackage{bm}

\begin{document}


\title{Surface energies, work functions, and surface relaxations of low index metallic surfaces from first-principles}

\author{Nicholas E. Singh-Miller}
\email{nedward@mit.edu}
\author{Nicola Marzari}

\affiliation{%
Department of Materials Science and Engineering, and Institute for Soldier Nanotechnologies, Massachusetts Institute of Technology, Cambridge, MA 02139
}%

\date{\today}

\begin{abstract}
We study the relaxations, surface energies, and work functions of low index metallic surfaces using pseudopotential plane-wave density-functional calculations within the generalized gradient approximation.  We study here the (100), (110), and (111) surfaces of Al, Pd, Pt, and Au and the (0001) surface of Ti, chosen for their use as contact or lead materials in nanoscale devices.  We consider clean, mostly non-reconstructed surfaces in the slab-supercell approximation.  Particular attention is paid to the convergence of these quantities with respect to slab thickness; furthermore, different methodologies for the calculation of work functions and surfaces energies are compared.  We find that the use of bulk references for calculations of surface energies and work functions can be detrimental to convergence unless numerical grids are closely matched, especially when surface relaxations are being considered.  Our results and comparison show that calculated values often do not quantitatively match experimental values.  This may be understandable for the surface relaxations and surface energies, where experimental values can have large error, but even for the work functions, neither local nor semi-local functionals emerge as an accurate choice for every case.  
\end{abstract}

\pacs{68.47.De, 71.17.Mb, 71.15.Nc}
\keywords{surface energy, work function, density functional theory, GGA}
\maketitle
\section{\label{sec:intro}Introduction}
Electronic structure calculations are becoming more widely applied to increasingly complex and realistic materials systems and devices, reaching well into the domain of nanotechnology with applications ranging from metal-molecule junctions, carbon-nanotube field effect transistors, and nanostructured metals or semiconductors.\cite{Hafner, Marzari2}  For such complex applications, characterizing the properties of the elementary building blocks becomes of fundamental importance.  Consider nanoscale constructions such as self assembled monolayers, where the underlying metal work function (and Fermi energy) are crucial to determining interfacial phenomena.\cite{Heimel}  Thus, it becomes critical to have a clear understanding of the fundamental properties of the metal lead surfaces -- starting from surface energies, structural relaxations, and work functions.  

It is also important to highlight the limitations of density functional theory when considering even only the building blocks, in this case with regard to surface properties of metals.  In general local and semi-local approximations for the exchange-correlation functional perform well for metals, however this is not the case for metal surfaces.\cite{Stroppa}  For this purpose we proceed here with a detailed study of the properties of clean metal surfaces of interest, as lead materials, to nanotechnological applications, and discuss the issues involved in obtaining converged estimates using electronic-structure modeling - in this case using density-functional theory (DFT) and the slab-supercell approximation - and the accuracy of these estimates compared to experimental values.

First, surface relaxations can arise from the creation of a new surface -- i.e. cleaving of a bulk in two.  This leads to a smoothing of the charge density at the new surface which causes a net force on the outermost surface layer of ions pointing into the bulk.\cite{Finnis}  DFT calculations in the slab-supercell approximation, assuming a pristine unit cell of the unreconstructed surface, can be effectively used to study these surface layer relaxations.

Second, the surface energy is the energy required to cleave an infinite crystal in two -- i.e. the amount of energy required to create a new surface.  This is a difficult quantity to determine experimentally because it usually requires measuring surface tension at the melting temperature of the metal.\cite{Tyson}  Theoretical determination of this quantity is relatively easy and particularly useful in studies of the relative stability of different surface facets.\cite{Stumpf}  However surface energy calculations within DFT are sensitive to numerical errors arising from differences in Brillouin-zone sampling.\cite{Boettger}  Methodologies for avoiding these errors have been proposed in the literature\cite{Boettger,Methfessel,DaSilva} and are discussed in later sections of this study. 

Third, the work function is the minimum energy needed to remove an electron from the bulk of a material through a surface to a point outside the material.  In practice, this is the energy required at $0$ K to remove an electron from the Fermi level of the metal to the vacuum potential.\cite{Lang}  Calculations of work function using DFT employ this definition and determine the Fermi energy and vacuum potential from calculations of the metals in slab-supercell geometries.  However, work functions calculated with slab approximations are known to have a dependency on the thickness of the slab, thus further analysis is required to extract bulk metal work functions from slab approximation.  This dependency is well documented in some cases and is attributed to finite size effects arising from classical electrostatic interactions or from quantum size effects (QSE).\cite{Feibelman2,Ciraci,Wei}  Methodology to lessen such effects is available in the literature.\cite{Fall}

In this paper we calculate the relaxations, surface energies, and work functions of the (111), (100), and (110) surfaces of Al, Pd, Pt, and Au and the (0001) surface of Ti within DFT.  We examine the convergence of all three quantities with respect to the thickness of the slabs, following the surfaces of Pd as an example.  Furthermore, we employ and compare a number of different methods of calculation with respect to surface energy and work function.  Well-converged values of these quantities for all of the metals and surfaces of this study are tabulated and compared, when available, to experimental quantities and/or DFT-LDA quantities from the literature.  Some of the LDA generated data from the literature, and compared to here, were calculated with surface Green's functions methodologies.\cite{Methfessel, Vitos, Benesh}  These methods have been shown to be highly accurate, as well as inherently free from finite-size effects due to choices of boundary conditions.  However, we feel that the flexibility and popularity of plane-wave DFT methods for more complex systems highlights the necessity for comparison of methods and functionals with well converged slab-supercell calculated quantities.

\section{\label{sec:methods}Methodology}
First-principles calculations within density functional theory are carried out using the PWscf code of the Quantum-ESPRESSO distribution.\cite{QE}  The Perdew-Burke-Ernzerhof (PBE) functional within the generalized gradient approximation (GGA) is used.\cite{Perdew3}  For the Brillouin-zone integration, we use a Monkhorst-Pack set of special {\bf k}-points,\cite{Monkhorst} and Marzari-Vanderbilt smearing \cite{Marzari} with a broadening of 0.02 Ryd. 


Ultrasoft pseudopotentials (USPP) are used for Pd, Pt, and Au.  For the case of Pd and Pt the USPPs used have been generated with the RRKJ3 scheme \cite{Rappe} with 10 valence electrons each in 4d$^9$5s$^1$ and 5d$^9$6s$^1$ configurations, respectively.  The Au USPP have also been generated with the RRKJ3 scheme with 11 valence electrons in a 5d$^{10}$6s$^1$ configuration.\footnote{We used the pseudopotentials Al.pbe-rrkj.UPF, Pd.pbe-nd-rrkjus.UPF, Pt.pbe-rrkjus.UPF, and Au.pbe-d-rrkjus.UPF from the http://www.quantum-espresso.org distribution.}  For the case of Ti the USPP was generated using the Vanderbilt scheme \cite{Vanderbilt} with 12 valence electrons in a 3s$^2$3p$^6$4s$^2$3d$^2$ configuration.  A normconserving pseudopotential is used for Al with electrons in a 3s$^{2}$3p$^1$ configuration.  The kinetic energy cutoff for the plane wave basis are 32 Ryd for the wave function and 512 Ryd for the charge density for Al, Pd, Pt, and Au.  For Ti these values are 40 Ryd and 640 Ryd, respectively. 

Surfaces were constructed using a supercell with a thin slab of metal separated from its periodic images by a layer of vacuum.  The size of this region is such that there are always $\sim 16$\AA\, of vacuum between the surfaces.  For all the slabs considered, each layer in the unit cell contains one in-equivalent atom.  For (111) surfaces a hexagonal cell with a base defined by $a_0[110]$ and $a_0[101]$ and an inter-layer spacing of $\frac{a_0}{\sqrt{3}}$ and a stacking of ABCABC is used, where $a_0$ is the equilibrium lattice parameter.  The other surfaces are similarly constructed; for (100) a tetragonal cell is used with a base formed by $\frac{a_0}{\sqrt{2}}[110]$ and $\frac{a_0}{\sqrt{2}}[1\bar{1}0]$ where the inter-layer spacing is $\frac{a_0}{2}$ and ABAB stacking.  The (110) surface is constructed from an orthorhombic cell with a base defined by $\frac{a_0}{\sqrt{2}}[110]$ and $a_0[100]$ with an inter-layer spacing of $\frac{a_0\sqrt{2}}{4}$ and ABAB stacking.  For the case of Ti the hexagonal cell is used with a ABAB stacking in the $z$-direction.

The remaining methodological items are discussed in the following results section.


\section{\label{sec:relax}Results}
\subsection{Bulk Properties}

We calculate the lattice parameters and bulk moduli for Al, Pd, Pt, Au, and Ti.  The calculations are performed for the total energy of the bulk system for a range of lattice parameters $a$.  The total energy data are fit with the Murnaghan equation of state\cite{Murnaghan} to obtain the bulk moduli.  For the case of bulk Ti, a hexagonal cell is used with a two atom basis to construct the hexagonal close packed (HCP) structure.  Total energy calculations are performed for a range of $c/a$ ratios and each of them fit with the Murnaghan equation of state.  The results are found in Table \ref{tab:lat}.  Here we see that the lattice parameters of the FCC Al, Pd, Pt, and Au are over estimated ($< 2\%$) by the use of the PBE-GGA exchange correlation functional.  Similarly, for the bulk moduli the calculated values are underestimated with respect to the experimental values.  However, for the calculations of these properties of Ti, PBE slightly overestimates the volume of the unit cell as well as over-estimating the bulk modulus.  This type of behavior can be seen in a study of similar metals when calculated using the PBE functional and a full potential linear augmented plane wave (FLAPW) all electron methodology.\cite{DaSilva}  From the same FLAPW study, we see good agreement with our pseudopotential calculations for the lattice parameters (within 0.03\AA) and the bulk moduli (within 5 GPA) of Al, Pd, and Pt.\cite{DaSilva}  The remainder of this work the equilibrium lattice parameters calculated with PBE, found in Table \ref{tab:lat}, are used.  

Recently in the literature, it has been shown that Pd calculated within GGA is incorrectly described as having a magnetic ground state in the bulk.\cite{Alexandre, Stewart}  To this end we also have calculated the bulk and surface properties of Pd with spin-polarization.  We also find that the ground state is described as magnetic, however, when applied to the surface properties, the introduction of magnetism has a small effect.  These calculations are reported and discussed in the Appendix of this paper.

\begin{table}
\caption{\label{tab:lat}Calculated lattice parameters and bulk moduli of the metals considered in this study compared with experimental values.}
\begin{ruledtabular}
\begin{tabular}{cccll}
   & $a_0$ (\AA)(c/a)& $a_{0,\text{expt.}}$ (\AA)(c/a) &B (GPa)& B$_{expt.}$(GPa)\\
\hline
Al& 4.06       & 4.05\footnotemark[1]       &74   & 79\footnotemark[1]\\
Pd& 3.98       & 3.89\footnotemark[1]       &163  & 195\footnotemark[2] \\
Pt& 3.99       & 3.92\footnotemark[1]       &246  & 288\footnotemark[3] \\
Au& 4.16       & 4.08\footnotemark[1]       &140  & 180\footnotemark[4] \\
Ti& 2.95 (1.57) & 2.95 (1.59)\footnotemark[1] &121 & 110\footnotemark[5]
\footnotetext[1]{Ref.\cite{Kittel}}
\footnotetext[2]{Ref.\cite{Rayne}}
\footnotetext[3]{Ref.\cite{MacFarlane}}
\footnotetext[4]{Ref.\cite{Neighbours}}
\footnotetext[5]{Ref.\cite{Fisher}}
\end{tabular}
\end{ruledtabular}
\end{table}


\subsection{Surface Relaxations}
Surface relaxation is characterized as the percent change $\Delta d_{ij}$ of the spacing between layers $i$ and $j$ versus the equilibrium layer spacing $d_0$.  For the (111), (100), (110), and (0001) surfaces $d_0$ is $a_0/\sqrt{3}$, $a_0/\sqrt{2}$, $a_0\sqrt{2}/4$, and $c_0/2$, respectively.
  
\begin{figure}
\includegraphics[width=3.2in]{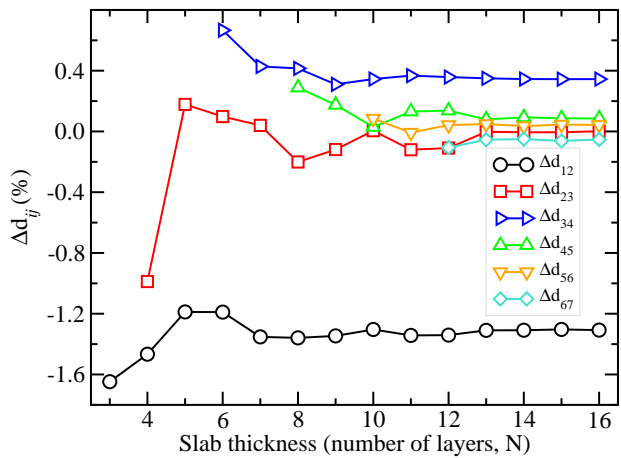}
\caption{\label{fig:relax}  Layer relaxations for the top six layers of Pd(100) as a function of slab thickness.  The numbers reported are the percent change, $\Delta d_{ij}$, of the spacing between layers $i$ and $j$ versus the initial layer spacing $d_0$ (for the (100) case $d_0=a_0/\sqrt{2}$).}
\end{figure}
As an example, the layer relaxations for Pd(100) as a function of slab thickness can be seen in Fig. \ref{fig:relax}:  Convergence of the relaxed layer spacing is achieved with increasing number of layers, and a slab thickness of 13-layers assures all relaxations are converged to within 0.1\%.  We show in Table \ref{tab:relax} the top three surface layer relaxations for all surfaces for 13-layer slabs.  As expected, the relaxation of the surface layers is related to the density of packing,\cite{Finnis} with larger relaxation for the less-densely packed surfaces, with patterns of multilayer relaxation that become noticeable as we go from (111) to (100) to (110) surfaces.  For the case of Ti(0001), the surface is hexagonal close-packed, however the large layer spacing leads to large relaxations.  Comparison to low-energy electron diffraction (LEED) and other experimental values shows that our well converged values do not correspond directly to the experimentally observed values.  However, regarding first layer relaxation or expansion, the qualitative trend is captured in all but the (100) surfaces.

Finally, we note that for this study we are typically concerned with clean non-reconstructed surfaces (larger supercells in the in-plane dimension could allow for surface reconstruction).  Some of the more widely know surface reconstructions are the $(22\times\sqrt{3})$-reconstruction of the Au(111) surface,\cite{Perdereau} the $(2\times 1)$ missing-row reconstruction of the (110) surfaces of Pt and Au,\cite{Wolf,Kellogg} and a number of reconstructions of the (100) surfaces of Au and Pt, the most studied being the (100)-hex.\cite{Fedak,Heilmann}  To this end we investigate for comparison the Au(110) and Pt(110)$(2\times 1)$-missing row reconstruction for 13-layer slabs.  We find that in both cases the first-layer relaxation more closely matches that of the experimental values in Table \ref{tab:relax}; $-19.7\%$ and $-18.6\%$ for Au and Pt, respectively.  However, no improvement is seen in the second layer relaxation.  Finally, the surfaces of the $4d$-metal Pd as well as the Ti(0001) surface are know to be less likely to reconstruct at low temperatures.  For the case of surface relaxations, neither the LDA nor GGA exchange-correlation functionals appear to represent a more accurate choice when compared to experimental values.

\begin{table*}
\caption{\label{tab:relax}  Surface relaxations for the top three layers of all surfaces considered in this study.  Reported are the values for a 13-layer slabs, compared to LDA and experimental values from the literature.  For surfaces that are known to reconstruct, experimental values and those calculated as reconstructed in this study are given in square brackets.}
\begin{ruledtabular}
\begin{tabular}{cc|lll|lll|lll}
   &       &         &$\Delta d_{12} (\%)$&   &        &$\Delta d_{23} (\%)$&&          &$\Delta d_{34} (\%)$&       \\
   &Surface& \text{PBE}&LDA &Expt.&PBE&LDA&Expt.&PBE&LDA& Expt.\\
\hline
Al& (111) & $+1.04$\footnotemark[1],$+1.35$\footnotemark[2]         & $+1.35$\footnotemark[2] &   $+1.7\pm0.3$\footnotemark[3]  & $-0.54$\footnotemark[1],$+0.54$\footnotemark[2]         & $+0.54$\footnotemark[2]                        & $+0.5\pm 0.7$\footnotemark[3]                                & $+0.19$\footnotemark[1],$+1.06$\footnotemark[2]      & $+1.04$\footnotemark[2]                       &           \\                     

&&&&   $+1.4\pm0.5$\footnotemark[4]  &          &                        &                                 &       &                       &           \\
  & (100) &  $+1.73$\footnotemark[1]        &  $+0.5$\footnotemark[5] &   $+2.0\pm0.8$\footnotemark[6]  &  $+0.47$\footnotemark[1]        &                        &  $+1.2\pm0.7$\footnotemark[6]   &  $-0.27$\footnotemark[1]     &                       &  \\
  & (110) &  $-5.59$\footnotemark[1]        &  $-6.9$\footnotemark[7] &   $-8.4$\footnotemark[8]                 &    $+2.20$\footnotemark[1]                              &     & $+4.9$\footnotemark[8]     &  $-1.29$\footnotemark[1]                                &       &                                         \\

Pd& (111) & $+0.25$\footnotemark[1],$-0.01$\footnotemark[2]  &$-0.22$\footnotemark[2]&	$+1.3\pm 1.3$\footnotemark[9]   &$-0.34$\footnotemark[1],$-0.41$\footnotemark[2]   &$-0.53$\footnotemark[2] & $-1.3\pm 1.3$\footnotemark[9]    &$+0.10$\footnotemark[1],$-0.22$\footnotemark[2]&$-0.33$\footnotemark[2]&	$+2.2\pm 1.3$\footnotemark[9]    \\
  &       &          &$-0.1$\footnotemark[10] & $+0.0\pm4.4$\footnotemark[11] &&&&&&\\
  & (100) & $-1.30$\footnotemark[1]  &$-0.6$\footnotemark[10]       &$+3.0\pm 1.5$\footnotemark[12]	   &$-0.00$\footnotemark[1]   &	   &$-1.0\pm 1.5$\footnotemark[12]	  &$+0.35$\footnotemark[1]&     &	    \\
  &       &          & &$+0.3\pm 2.6$\footnotemark[13]	   &  &	   &	  &&     &	    \\  
  & (110) & $-8.49$\footnotemark[1]  &$-5.3$\footnotemark[10]	   &$-5.8\pm 2.2$\footnotemark[14]	   &$+3.47$\footnotemark[1]   &	   &$+1.0\pm 2.2$\footnotemark[14]	  &$-0.19$\footnotemark[1]&     &	    \\
  &  &   &	   &$-5.1\pm 1.5$\footnotemark[15] &                                       &	      &	$+2.9\pm 1.5$\footnotemark[15]   &&     &	    \\
Pt& (111) & $+0.85$\footnotemark[1],$+1.14$\footnotemark[2]  &$+0.88$\footnotemark[2]&$+1.1\pm 4.4$\footnotemark[16]	   &$-0.56$\footnotemark[1],$-0.29$\footnotemark[2]&$-0.22$\footnotemark[2]&	  &$-0.15$\footnotemark[1],$-0.21$\footnotemark[2]&$-0.17$\footnotemark[2]     &	    \\
&& &&$+0.5\pm 0.9$\footnotemark[17]	   &&&	  &     &	    \\
&& &&$+1.4\pm 0.9$\footnotemark[18]	   &&&	  &     &	    \\
  & (100) & $-2.37$\footnotemark[1]  &	   &$+0.2\pm 2.6$\footnotemark[19]	   &$-0.55$\footnotemark[1]  &	   &	  &$+0.29$\footnotemark[1]&     &	    \\
  & (110) & $-15.03$\footnotemark[1] &	   &[$-18.5\pm 2.2$]\footnotemark[20]	   &$+7.61$\footnotemark[1]   &	   &[$-24.2\pm 4.3$]\footnotemark[20]	  &$-1.70$\footnotemark[1]&     &	    \\
  &       & [$-18.62$]\footnotemark[1]         &	   &[$-19.5\pm 7.2$]\footnotemark[21]	   & [$+10.59$]\footnotemark[1]   &	   &[$-7.9\pm 5.8$]\footnotemark[21]	  & [$-9.57$]\footnotemark[1]&     &	    \\

Au& (111) & $-0.04$\footnotemark[1]  &$+0.8$\footnotemark[22]		     	   &			   	   &$-1.86$\footnotemark[1]   &$-0.3$\footnotemark[22]			     	   &	  &$-1.40$&     &	    \\
  & (100) & $-1.51$\footnotemark[1]  &	$-1.2$\footnotemark[23]		     	   &[$-20\pm 3$]\footnotemark[24]			   	   &$+0.33$\footnotemark[1]   &	$+0.4$\footnotemark[23]		     	   &[$+2\pm 3$]\footnotemark[24]			  	  &$+0.24$\footnotemark[1]&     &	    \\

  & (110) & $-12.94$\footnotemark[1] &	$-9.8$\footnotemark[25]		     	   &[$-20.1\pm 3.5$]\footnotemark[26]			   	   &$+7.83$\footnotemark[1]   &	$+7.8$\footnotemark[19]&[$-6.2\pm 3.5$]\footnotemark[26]	&$-2.66$\footnotemark[1]&	$-0.8$\footnotemark[25] &	    \\
  & &[$-19.7$]\footnotemark[1]  &	&[$-18.1\pm 6.9$]\footnotemark[27] &[$+10.45$]\footnotemark[1]  &&[$-6.8\pm 6.9$]\footnotemark[27]	&[$-11.03$]\footnotemark[1]&&	    \\
  & &  &	&[$-22.2\pm 6.9$]\footnotemark[21] &  &&	&&&	    \\

Ti& (0001)& $-6.47$\footnotemark[1],$-6.84$\footnotemark[2]  &$-6.44$\footnotemark[2]  &-4.9\footnotemark[28]	   &$+4.01$\footnotemark[1],$+2.82$\footnotemark[2]   &$+2.64$\footnotemark[2] &$+1.4$\footnotemark[28]	  &$-0.68$\footnotemark[1],$-0.51$\footnotemark[2]&$+0.37$\footnotemark[2]&$-1.1$\footnotemark[28]
\footnotetext[1]{Present study}
\footnotetext[2]{FLAPW-LDA-Slab \cite{DaSilva}}
\footnotetext[3]{LEED \cite{Noonan}}
\footnotetext[4]{LEED \cite{Burchhardt}}
\footnotetext[5]{PWPP-LDA-Slab \cite{Borg}}
\footnotetext[6]{LEED \cite{Petersen}}
\footnotetext[7]{PWPP-LDA-Slab \cite{Kiejna2}}
\footnotetext[8]{PWPP-LDA-Slab \cite{Schochlin}}
\footnotetext[9]{LEED \cite{Ohtani}}
\footnotetext[10]{LDA-SGF \cite{Methfessel}}
\footnotetext[11]{HEIS \cite{Kuk}}
\footnotetext[12]{LEED \cite{Quinn}}
\footnotetext[13]{LEED \cite{Behm}}
\footnotetext[14]{LEED \cite{Barnes}}
\footnotetext[15]{LEED \cite{Skottke}}
\footnotetext[16]{LEED \cite{Adams}}
\footnotetext[17]{SPLEED \cite{Feder}}
\footnotetext[18]{MEIS \cite{VanDerVeen}}
\footnotetext[19]{HEIS \cite{Davies}}
\footnotetext[20]{$(1\times 2)$ LEED \cite{Sowa}}
\footnotetext[21]{$(1\times 2)$ XRD \cite{Vlieg}}
\footnotetext[22]{mixed basis-PP-LDA-Slab\cite{Takeuchi}}
\footnotetext[23]{PWPP-LDA-Slab\cite{Yu2}}
\footnotetext[24]{\emph{hex} XRD\cite{Ocko}}
\footnotetext[25]{mixed basis-PP-LDA-Slab\cite{Bohnen}}
\footnotetext[26]{$(1\times 2)$ LEED \cite{Moritz}}
\footnotetext[27]{$(1\times 2)$ MEIS \cite{Copel}}
\footnotetext[28]{LEED \cite{Teeter}}
\end{tabular}
\end{ruledtabular}
\end{table*}

\subsection{\label{sec:surf}Surface Energies}
\begin{figure}
\includegraphics[width=3.2in]{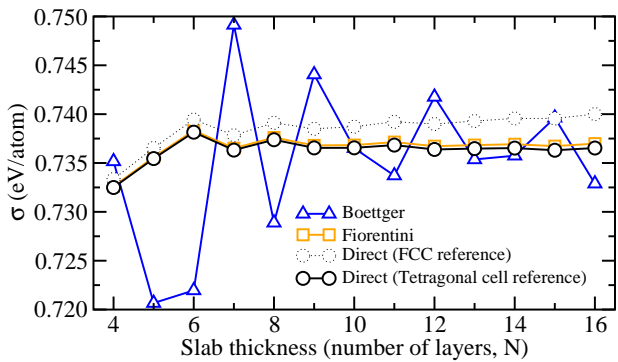}
\caption{\label{fig:surf} Surface energy versus slab thickness for the Pd(100) surface calculated with the methods of Boettger\cite{Boettger}, Fiorentini and Methfessel\cite{Fiorentini}, and directly from Eq. \ref{eqn:surf} with two different values for $E_{\textrm{bulk}}$.}
\end{figure}

The surface energy is the energy required to create a new surface, and as mentioned earlier, it is a difficult quantity to determine experimentally.  In our calculations the surface energy $\sigma$ can be determined by taking the energy difference between the total energy of a slab and an equivalent bulk reference amount, as seen in the following expression:
\begin{equation}\label{eqn:surf}
\sigma=\lim_{N\to\infty}\frac{1}{2}(E_{\textrm{slab}}^N-NE_{\textrm{bulk}}),
\end{equation}
where $E_{\textrm{slab}}^N$ is the total energy of an N-atom slab, $E_{\textrm{bulk}}$ is the total energy of the bulk per atom, and the factor $\frac{1}{2}$ accounts for the two surfaces in the slab unit-cell.  However, this expression will diverge with increasing slab thickness if there are numerical differences between the calculation for the bulk and the slab (such as differences in the {\bf k}-point mesh, etc.).\cite{Boettger}  Two methods have been suggested to determine surface energies while avoiding this divergence.  Boettger\cite{Boettger} uses the bulk energy in Eq. \ref{eqn:surf} as $E_{\textrm{slab}}^N- E_{\textrm{slab}}^{N-1}$, thus avoiding a calculation on a separate bulk system and effectively eliminating the errors from differences in {\bf k}-point sampling, while Fiorentini and Methfessel\cite{Fiorentini} make the assumption that in the limit of large $N$ one can rewrite Eq. \ref{eqn:surf} as:
\begin{equation}\label{eqn:linear}
E_{\textrm{slab}}^N\approx 2\sigma+NE_{\textrm{bulk}}.
\end{equation}
If the total energy of the slab depends linearly on slab thickness $N$, the bulk energy term $E_{\textrm{bulk}}$ can be taken as the slope and used in Eq. \ref{eqn:surf}, also avoiding a calculation on a separate bulk system.  In practice, it has been shown that divergence can be avoided when large and matching {\bf k}-point samplings are used for the slab and bulk calculations.\cite{DaSilva}  Here we compare these three methods using the labels ``Boettger'', ``Fiorentini'', and ``direct'' respectively.  Fig. \ref{fig:surf} shows the comparison of surface energy versus slab thickness for Pd(100) surface energies.  Here two different calculations are performed for the ``direct'' determination of $E_{\textrm{bulk}}$.  First, a single FCC unit cell with a $16\times16\times16$ {\bf k}-point mesh (dashed circles in Fig. \ref{fig:surf}) and second a two atom tetragonal unit cell with a $16\times16\times11$ k-point mesh to, as closely as possible,  match that of the {\bf k}-point mesh used for the slab (solid circles in Fig. \ref{fig:surf}).   

\begin{figure}
\includegraphics[width=3.2in]{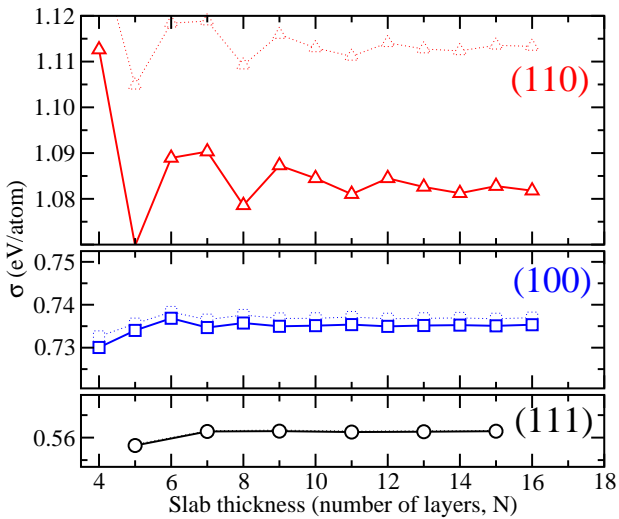}
\caption{\label{fig:surfPd} Surface energies of unrelaxed (dashed lines) and relaxed (solid lines) slabs of Pd(110), Pd(100), and Pd(111) surfaces versus slab thickness.}
\end{figure}

It can be seen from Fig. \ref{fig:surf} that the best non-divergent results as a function of slab thickness are obtained with the method of Fiorentini and Methfessel, while the method of Boettger oscillates around these converged values.  This periodicity is likely a manifestation of the finite size effects discussed earlier.  Furthermore, even though a linear regression of the total energy versus slab thickness yields an excellent fit, the localized slopes ($E_{\textrm{slab}}^N- E_{\textrm{slab}}^{N-1}$) of the Boettger method lead to  energy fluctuations in the bulk reference energy of the order $1\times 10^{-3}$eV.  These fluctuations are further amplified by the factor $-N$.  These same fluctuations can be seen in all methods when comparing surface energy data for any given slab thicknesses $N$ and $N\pm 1$; however, the choice of bulk reference in the direct and Fiorentini methods do not further amplify them.  Finally, neither the Boettger or Fiorentini methods diverge in the same manner that the direct method does when the {\bf k}-point grids do not match well for the bulk and surface calculations (dotted circles in Fig. \ref{fig:surf}).  However, similar to the findings of Da Silva \emph{et al.},\cite{DaSilva} we find that if the same care is used when choosing the {\bf k}-point mesh (as is the case for the solid circles in Fig. \ref{fig:surf}) an acceptable result is achievable through the direct method.  For the remainder of this paper reported surface energies are calculated with the method of Fiorentini and Methfessel. 

\begin{table}
\caption{\label{tab:seng} Surface energies for 13-layer slabs in both unrelaxed and fully relaxed geometries reported here in (eV/atom).  LDA and experimental values compared with the surface energy of the relaxed surfaces reported here in units of (J/m$^2$).}
\begin{ruledtabular}
\begin{tabular}{crcclcc}
   &Surface&$\sigma_u$\footnotemark[1]&$\sigma_r$\footnotemark[1]&$\gamma_{\text{PBE}}$&$\gamma_{\text{LDA}}$&$\gamma_{\text{Expt.}}$\\
   &       & (eV/atom) & (eV/atom)&(J/m$^2$)&(J/m$^2$)&(J/m$^2$)\\
\hline
Al& (111) & 0.30 & 0.30 & 0.67\footnotemark[1],0.75\footnotemark[3] &0.91\footnotemark[3] & 1.14\footnotemark[2]\\
  & (100) & 0.45 & 0.44 & 0.86\footnotemark[1] &			 &\\
  & (110) & 0.70 & 0.68 & 0.93\footnotemark[1] &			 &\\
Pd& (111) & 0.56 & 0.56 &1.31\footnotemark[1],1.33\footnotemark[3]&1.87\footnotemark[3]&2.00\footnotemark[2]\\
  & (100) & 0.74 & 0.74 &1.49\footnotemark[1]&			 &\\
  & (110) & 1.11 & 1.08 &1.55\footnotemark[1]&			 &\\
Pt& (111) & 0.65 & 0.65 &1.49\footnotemark[1],1.67\footnotemark[3]&2.23\footnotemark[3]&2.49\footnotemark[2]\\
  & (100) & 0.91 & 0.90 &1.81\footnotemark[1]&			 &\\
  & (110) & 1.38 & 1.30 &1.85\footnotemark[1]&2.48\footnotemark[4]			 &\\
Au& (111) & 0.35 & 0.35 &0.74\footnotemark[1]&1.04\footnotemark[5]&1.50\footnotemark[2]\\
  & (100) & 0.46 & 0.46 &0.85\footnotemark[1]&1.39\footnotemark[6]			 &\\
  & (110) & 0.71 & 0.69 &0.90\footnotemark[1]&1.55\footnotemark[4]			 &\\
Ti& (0001)& 0.97 & 0.92 &1.96\footnotemark[1],1.99\footnotemark[3]&2.27\footnotemark[3]&1.99\footnotemark[2]

\footnotetext[1]{Present study}
\footnotetext[2]{Ref.\cite{Tyson}}
\footnotetext[3]{FLAPW-LDA-Slab\cite{DaSilva}}
\footnotetext[4]{PWPP-LDA-Slab\cite{Lozovoi2}}
\footnotetext[5]{mixed basis-PP-LDA-Slab\cite{Takeuchi}}
\footnotetext[6]{PWPP-LDA-Slab\cite{Yu2}}
\end{tabular}
\end{ruledtabular}
\end{table}

Surface energies are calculated for both relaxed (but unreconstructed) and unrelaxed surfaces of all metals considered in this study. As an example, Fig. \ref{fig:surfPd} shows the surface energies as a function of slab thickness for all three Pd surfaces, both unrelaxed (dashed lines) and relaxed (solid lines).  From the figure it is clear that surface energies are well converged for slabs as thin as 6-layers.  Also, as expected, the surfaces that relax the most -- i.e. (110) -- experience the largest change in surface energy.  

Calculated values for 13-layer slabs for all metals studied can be found in Table \ref{tab:seng}.  All of the values calculated are converged with an accuracy comparable to those of Fig. \ref{fig:surfPd}.  Furthermore, in the table we compare our present PBE calculated values with LDA and experimental values from the literature.  Remarkably, PBE seems to underestimate surface energies by as much as a factor of 2.  If better comparison to experimental values is desired, LDA represents the better choice of exchange-correlation functional for all of the (111) surfaces for which data was available.  However, the converged value of surface energy for the Ti(0001) surface, calculated within PBE-GGA appears to more closely match the experimental data.  Furthermore, regarding the (111) surfaces, our psedopotential calculations show excellent agreement with the FLAPW calculations\cite{DaSilva} having only small differences all less than 0.03 eV/atom for the metals studied.

In the context of poorly represented surface energies, we note the existence of exchange-correlation functionals that are designed to address this class of problem.  First, one in which the surface and bulk of the material are treated with region specific functionals with an interpolation region between them.\cite{Mattsson,Armiento}  As Armiento and Mattsson point out, the functional they develop in their present article (AM05) give results similar to LDA, and while they have not been used on molecules, they represent a good choice for solid state materials.\cite{Armiento}  Second, in removing the bias toward the description of free-atom energies through a restoration of the second-order gradient expansion for exchange over a wide range of densities Perdew \emph{et al.} create a functional built on PBE, PBEsol.\cite{Perdew7}  This same construction principle has more recently been applied to the meta-GGA functional (TPSS) by Perdew \emph{et al.} and also shows promising results.\cite{Perdew6}  The promise of these functionals (AM05 and PBEsol) regarding bulk properties is further supported by a recent overview of the accuracy of new functionals for bulk solids by Csonka \emph{et al.}\cite{Csonka}  It is shown that both functionals perform notably better for most metals.  Furthermore, regarding bulk Pd, which in our study shows the largest discrepancy with respect to experiment, they show an agreement of the lattice constant within 0.003 or less for both functionals.\cite{Csonka}  Ropo \emph{et al.} compare the two functionals for calculation of surface energies of metallic surfaces, including the Pd(111) surface, where they show an agreement with the experimental value of 2.00 J/m$^2$ for both PBEsol (2.08 J/m$^2$) and AM05 (2.02 J/m$^2$).\cite{Ropo}  Further study of surfaces with these functionals are still needed, and the popularity of the functionals PBE and LDA still dominates the study of extended systems.

\subsection{\label{sec:wf}Work Function}
As mentioned earlier, the work function is the minimum energy needed to remove an electron from the bulk of a material through a surface to a point outside the material, and can be written as:
\begin{equation}\label{eqn:wf}
\Phi=V_{\textrm{vacuum}}-E_F.
\end{equation}
Again we need to test the calculation of the work function through use of the slab-supercell approximation.  Thus, we compare and discuss the calculation of work functions \emph{a posteriori} by two methods.  First, we use directly Eq. \ref{eqn:wf}, where both the potential in the vacuum region $V_{\textrm{vacuum}}$ and the Fermi energy $E_F$ are derived from the same calculation.  Second, we use a methodology utilizing macroscopic averages.\cite{Balderschi, Fall} Here we take the macroscopic average of a potential step across the surface of the slab, $\Delta V=V_{\textrm{vacuum}}-V_{\textrm{slab interior}}$, and reference it to the macroscopic average of a potential obtained from a calculation on a separate bulk system, $V_{\textrm{bulk}}^{\textrm{Macro}}$.  This allows for the use of the Fermi energy from the bulk system $E_{F,\textrm{bulk}}$, which will not suffer from quantum size effects.\cite{Fall}
\begin{equation}\label{eqn:macro}
\Phi=\Delta V+V_{\textrm{bulk}}^{\textrm{Macro}}-E_{F,\textrm{bulk}}
\end{equation} 
Note here that the window for macroscopic averaging is the unrelaxed equilibrium layer spacing.  All of the potentials discussed here (both from the slab and the bulk reference) refer to the electrostatic part of the total potential (the Hartree potential).  This part of the potential tails off more rapidly in the vacuum region of the slab-supercell when compared to the full Kohn-Sham potential including the exchange-correlation potential.  

\begin{figure}
\includegraphics[width=3.2in]{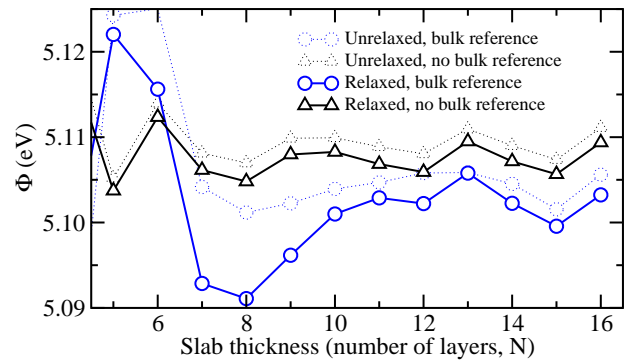}
\caption{\label{fig:wfcomp} A comparison of work function calculations methods for work functions of Pd(100) slabs versus slab thickness.  The bulk reference used is a single unit cell of FCC Pd.}
\end{figure}

A comparison of these methods of calculation can be seen in Fig. \ref{fig:wfcomp}.  Here the work function of Pd(100) slabs is plotted versus slab thickness for both the unrelaxed slabs (dashed lines) and relaxed slabs (solid lines).  It is clear from the figure that both methods converge quickly for the unrelaxed case (in as little as 7-layers).  However, when taking relaxation of all layers into account, the method of macroscopic averaging converges much more slowly.  This is due to the fact that the averaging window for the reference potential and the layer spacing of the interior of the slab are not equivalent.  Work functions reported in the remainder of this paper are taken directly by Eq. \ref{eqn:wf} without bulk references.  

\begin{figure}
\includegraphics[width=3.2in]{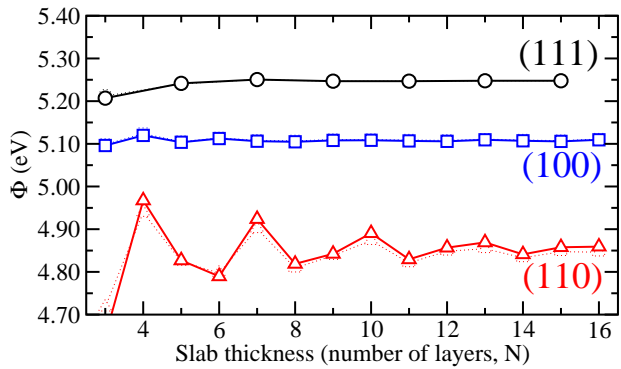}
\caption{\label{fig:wf} Work function versus slab thickness for the (111), (100), and (110) Pd surfaces, both unrelaxed (dashed lines) and relaxed (solid lines).}
\end{figure}

As a demonstration of the convergence of the work function as a function of slab thickness, the results for three Pd surfaces are shown in Fig. \ref{fig:wf}.  Here it can be seen that the work function of the (111) and (100) surfaces converges much more quickly than for the (110) surface.  Even so, the work function of the (110) surface is converged to within $0.1$eV by 8-layers and $< 0.05$eV by 11-layers.  As in the previous subsection regarding the surface energy, oscillations can be seen.  Furthermore, it is know that the work function of different facets can differ greatly\cite{Fall2} and the general trend is that work function will decrease with decreasing layer packing, which is consistent with the findings in Fig. \ref{fig:wf}.

Our final results for the work function of all the metals considered can be seen in Table \ref{tab:wf}; LDA and experimental values from the literature are provided when available.  With the exception of Au and Ti, LDA again represents the better choice of exchange-correlation functional if a better match to experimental results is desired.  We find it particular interesting that Au and Pt should behave so differently with respect to the functional used.  Regarding Al, we see the same trend as reported in Da Silva \emph{et al.}, where the work function of the (111) and (110) surfaces are nearly equal, and the (100) has the larger of the three.  Also, recall that the (100) and (110) surfaces of Au and Pt are likely in a reconstructed state for the experimental values.  Finally, regarding the (111) surfaces, our psedopotential calculations again show excellent agreement with the FLAPW calculations\cite{DaSilva} having only small differences all less than 0.03 eV for the metals studied.

\begin{table}
\caption{\label{tab:wf}Work functions calculated for 13-layer slabs of all metal faces considered in this study.  The values in square braces are for surfaces that are likely reconstructed.  All values in eV.}
\begin{ruledtabular}
\begin{tabular}{cclll}
  & Surface &$\Phi$&$\Phi_{\text{LDA}}$ & $\Phi_{\text{Expt.}}$\\
\hline
Al & (111) & 4.02\footnotemark[1],4.04\footnotemark[4] &4.25\footnotemark[2],4.21\footnotemark[4]                       &4.23$\pm$0.02\footnotemark[3] \\
   & (100) & 4.30\footnotemark[1] &4.38\footnotemark[2]                       &4.42$\pm$0.03\footnotemark[3]	     \\
   & (110) & 4.09\footnotemark[1] &4.30\footnotemark[2]                       &4.12$\pm$0.02\footnotemark[3]	     \\
Pd & (111) & 5.25\footnotemark[1],5.22\footnotemark[4] &5.64\footnotemark[4]                        &5.90$\pm$0.01\footnotemark[13],5.55$\pm$0.01\footnotemark[5] \\
   & (100) & 5.11\footnotemark[1] &  	              &5.65$\pm$0.01\footnotemark[13]	     \\
   & (110) & 4.87\footnotemark[1] &	  	              &5.20$\pm$0.01\footnotemark[13]	     \\
Pt & (111) & 5.69\footnotemark[1],5.69\footnotemark[4] &6.06\footnotemark[4]                       &6.08$\pm$0.15\footnotemark[11],6.10$\pm$0.06\footnotemark[6]\\
   & (100) & 5.66\footnotemark[1] &	  	                              &[5.82$\pm$0.15\footnotemark[11]]	     \\
   & (110) & 5.26\footnotemark[1] &5.52\footnotemark[7]	  	              &[5.35$\pm$0.05\footnotemark[12]]	     \\
Au & (111) & 5.15\footnotemark[1] &5.63\footnotemark[8]	  	      &5.26$\pm$0.04\footnotemark[9]\\
   & (100) & 5.10\footnotemark[1] &5.53\footnotemark[8]	  	      &[5.22$\pm$0.04\footnotemark[9]]\\
   & (110) & 5.04\footnotemark[1] &5.41\footnotemark[8], 5.39\footnotemark[7]&[5.20$\pm$0.04\footnotemark[9]]\\
Ti & (0001)& 4.38\footnotemark[1],4.40\footnotemark[4] &4.66\footnotemark[4]                       & 4.33\footnotemark[10]
\footnotetext[1]{Present study}
\footnotetext[2]{PWPP-LDA-Slab\cite{Fall5}}
\footnotetext[3]{Ref.\cite{Grepstad}}
\footnotetext[4]{FLAPW-LDA-Slab\cite{DaSilva}}
\footnotetext[5]{Ref.\cite{Kubiak}}
\footnotetext[6]{Ref.\cite{Derry}}
\footnotetext[7]{PWPP-LDA-Slab\cite{Lozovoi2}}
\footnotetext[8]{PWPP-LDA-Slab\cite{Fall3}}
\footnotetext[9]{Ref.\cite{Hansson}}
\footnotetext[10]{Ref.\cite{Eastman}}
\footnotetext[11]{Ref.\cite{Salmeron}}
\footnotetext[12]{Ref.\cite{Vanselow}}
\footnotetext[13]{Ref.\cite{Hulse}}

\end{tabular}
\end{ruledtabular}
\end{table}


\section{\label{sec:summary}Summary}
Calculations of the relaxations, surface energies, and work functions of low index metallic surfaces were made for the (100), (110), and (111) surfaces of Al, Pd, Pt, and Au and the (0001) surface of Ti using finite slab approximations.  We have paid particular attention to the issues that arise from slab thickness and consequently finite size effects, and we have shown that convergence of these quantities can be achieved for slab thicknesses that are greater than 6-layers for (111),(100), and (0001) surfaces and 10-layers for (110) surfaces.  We find that the use of bulk references for calculations of surface energies and work functions can be detrimental to convergence, especially when surface relaxations are being considered.  

We compared our results within the GGA with experimental and LDA results from the literature.  Our results and comparison have shown that, even though converged values can be achieved, calculated values do not match well quantitatively to experimental values.  This may be understandable for the surface relaxations and surface energies, where even the experimental values have some room for interpretation.  While for the work functions, a property for which the experimental values are more reliable, we have found that neither LDA or GGA is unanimously a better choice.

\begin{acknowledgments}
This research has been supported by the Singapore-MIT Alliance, MIT Institute for Soldier Nanotechnologies (ISN-ARO No. DAAD 19-02-D-0002); computational facilities have been provided through NSF (No. DMR-0414849).
\end{acknowledgments}

\appendix*
\section{Magnetic Bulk Palladium}
\begin{figure}
\includegraphics[width=3.2in]{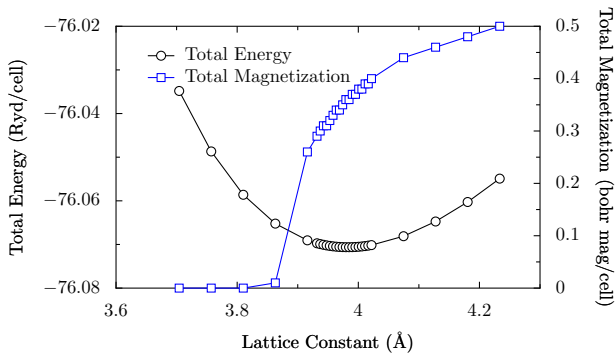}
\caption{\label{fig:Pdm1} Total energy per unit cell and total magnetization per unit cell plotted versus the FCC lattice constants of bulk Pd using the GGA-PBE exchange-correlation functional with spin-polarization.}
\end{figure}

\begin{figure}
\includegraphics[width=3.2in]{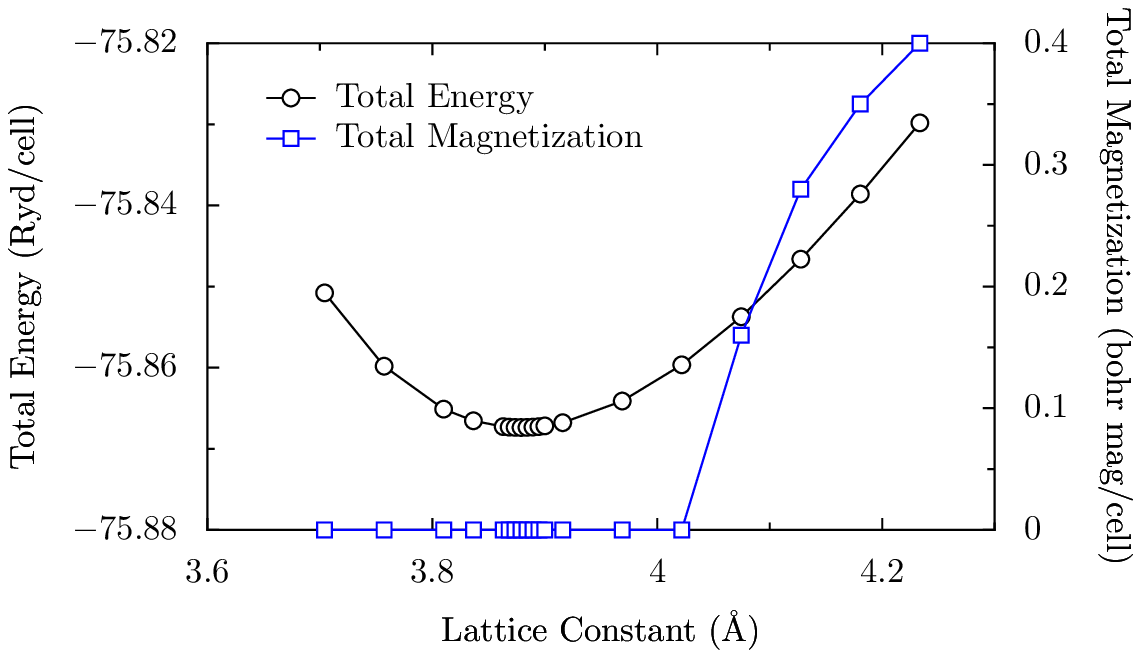}
\caption{\label{fig:Pdm2} Total energy per unit cell and total magnetization per unit cell plotted versus the FCC lattice constants of bulk Pd using the LDA exchange-correlation functional with spin-polarization.}
\end{figure}

Recent calculations in the literature have shown that when evaluated within the generalized gradient approximation (GGA) Pd yields an erroneous magnetic bulk ground state.\cite{Alexandre, Stewart}  In light of these observations we have performed a similar evaluation of the ground state of Pd using PBE-GGA and the pseudopotential mentioned in the text of this paper.  In a similar manner to Alexandre \emph{et al.} we find also a magnetic ground state for bulk Pd.  We plot both the total energy and the total magnetization as a function of FCC lattice parameter for Pd evaluated with GGA in Fig \ref{fig:Pdm1} and for comparison the same plot is generated for Pd within LDA in Fig \ref{fig:Pdm2}.

Finally, we have calculated the surface properties of the Pd(100) with spin-polarization.  In all cases the slab maintains a total magnetization, however, this has little effect on the converged surface properties discussed in the body of this paper.  The changes in the relaxations and surfaces energies are negligible, while the change in the work functions is seen as an increase of ~0.5\%.  


\end{document}